\documentclass[12pt]{article}

\usepackage{fancyhdr}
\usepackage{lastpage}
\usepackage[pdftex]{graphicx}
\usepackage{acronym}
\usepackage{amsmath} 
\usepackage{subfigure}
\newcommand{\comments}[1]{}

\newcommand{\ie}{\emph{i.e.}} 

\newcommand{\pa}{\emph{p.a.}}

\acrodef{OPEX}[OPEX]{Operating Expenses}
\acrodef{UE}[UE]{User Equipment}
\acrodef{BS}[BS]{Base Station}
\acrodef{DTX}[DTX]{Discontinuous Transmission}
\acrodef{PAPR}[PAPR]{Peak-to-Average Power Ratio }
\acrodef{SC-FDMA}[SC-FDMA]{Single-carrier FDMA}
\acrodef{FDMA}[FDMA]{Frequency Division Multiple Access}
\acrodef{TDMA}[TDMA]{Time Division Multiple Access}
\acrodef{CDMA}[CDMA]{Code Division Multiple Access}
\acrodef{ICT}[ICT]{Information and Communication Technologies}
\acrodef{QoS}[QoS]{Quality of Service}
\acrodef{PA}[PA]{Power Amplifier}
\acrodef{RS}[RS]{Resource Sharing}
\acrodef{PC}[PC]{Power Control}
\acrodef{SOTA}[SOTA]{State-Of-The-Art}
\acrodef{EE}[EE]{Energy Efficiency}
\acrodef{SINR}[SINR]{Signal-to-Interference-and-Noise-Ratio}
\acrodef{LTE}[LTE]{Long Term Evolution}
\acrodef{EARTH}[EARTH]{Energy Aware Radio and neTwork tecHnologies}

\begin{document}
\title{Fundamental Limits of Energy-Efficient Resource Sharing, Power Control and Discontinuous Transmission}

\author{ 
Hauke HOLTKAMP$^1$, Gunther AUER$^1$\\
$^1$DOCOMO Euro-Labs, Landsbergerstr. 312, \\
Munich 80687, Germany,\\
Email: \{lastname\}@docomolab-euro.com\\
}

\maketitle

\begin{abstract}
The achievable gains via power-optimal scheduling are investigated. Under the \acs{QoS} constraint of a guaranteed link rate, the overall power consumed by a cellular \ac{BS} is minimized. Available alternatives for the minimization of transmit power consumption are presented. The transmit power is derived for the two-user downlink situation. The analysis is extended to incorporate a \ac{BS} power model (which maps transmit power to supply power consumption) and the use of \ac{DTX} in a \ac{BS}. Overall potential gains are evaluated by comparison of a conventional \ac{SOTA} \ac{BS} with one that employs \ac{DTX} exclusively, a power control scheme and an optimal combined \ac{DTX} and power control scheme. Fundamental limits of the achievable savings are found to be at 5.5\,dB under low load and 2\,dB under high load when comparing the \ac{SOTA} consumption with optimal allocation under the chosen power model.
\end{abstract}


\section{Introduction}
Today, the battle against global warming -- pushed by recent initiatives -- requires energy optimization also on the \ac{BS} side -- in addition to the battery-limited \ac{UE} side. A decrease in total $\rm{CO_2}$ emission by 20\% by 2020 is requested \cite{bzb1001}. In a thriving sector like the \ac{ICT}, which is predicted to grow at 24\% percent \pa, this is a challenging goal. Globally, the \ac{ICT} sector causes 2\% of $\rm{CO_2}$ in the year 2007. Mobile communication accounts for a fraction of 0.3\% of global $\rm{CO_2}$. Achieving a 20\% decrease in emission by 2020 while growing in volume by 200\% necessitates a decrease in total energy consumption of today's mobile networks by 74\%. In addition to the reduction of $\rm{CO_2}$ emission, a lowering of \ac{OPEX} provides a strong incentive for network operators to reduce \ac{BS} energy consumption.

Over the evolution of cellular networks, the focus of research and development has been on maximizing throughput, spectral efficiency, reliability and \ac{QoS}. It has always been of utmost importance to provide the mobile user with a seamless, fast and dependable connection. While these aspects have not lost relevance, a new consideration is coming into play -- the energy efficiency of network operation. Due to the battery-powered nature of the mobile device, care has always been taken to minimize processing load on the \ac{UE} by passing it asymmetrically to the \ac{BS} which is usually mains powered. Examples for this design in \ac{LTE} include the use of OFDMA with a high \ac{PAPR} for downlink transmission (requiring more complex amplifiers) and \ac{SC-FDMA} on the uplink (placing complexity on the receiver side) and the BS always-on protocol design. In fact, operation of the mobile device over its life-time is so efficient today that it causes negligible power consumption compared to its manufacturing \cite{aghimfbhz1001}. While this asymmetry has achieved long battery lives on the mobile equipment side, little attention has been paid to the the overall network power consumption. Studies have shown that the power consumption of the infrastructure network per subscriber is between 10 and 100 times higher than the power consumption of the mobile device \cite{eon0801, fmbf1001}. In addition to this power consumption asymmetry, high spectrum efficiency is only required during peak traffic hours while during off-peak hours large parts of the available resources are unused. Fully exploiting these idle resources (\ie, transmit power, time and/or frequency) can greatly increase system efficiency \cite{czbfjgav1001}.

While the routine refinement of components has regularly been able to reduce energy consumption over multiple generations of devices \cite{bzb1001}, little work has been done on the energy efficiency of radio transmission. Often, these are just side mentions in throughput oriented research \cite{ssha0801a, ggok0601}. Furthermore, minimization of transmit power does not always lead to a minimization of supply (mains) power. The \ac{EARTH} project is based upon the above and has established the following mission: Reduce power consumption of the world-wide mobile networks by 50\% by improving the radio transceiver \ac{BS}, hereby focusing on \ac{LTE} and \ac{LTE}-Advanced as the upcoming mobile network. This work is concerned with effects of \ac{DTX} and power control on power savings while upholding a required link rate. Taking into account the supply power in addition to the transmit power, it is studied which approach uses the least energy. 

In this paper, we present the results of a study summarizing the fundamental limits in energy-efficient scheduling. Section~2 provides the necessary background for this analysis. In Section~3 the problem of minimizing transmit power is formally described. This model is extended in Section~4 to incorporate overall power consumption. We present the results and conclude in Section~5.

\section{Considerations for Energy-efficient Radio Resource Management}
\label{considerations}
When inspecting the internal power dissipation of a typical macro \ac{BS}, it turns out that around 65\% of the power is lost in the power amplifier \cite{fbzfgjt1001}. Power consumption of the power amplifier can be reduced in two ways. On the one hand, the efficiency of the component can be improved and thus the consumption reduced. On the other hand, the required transmit power can be reduced via appropriate scheduling. This study focuses on the latter. A more efficient and linear power amplifier will scale its consumption more directly with transmit power, thus both approaches support each other. \ac{BS} transmit power is reduced  via downlink \ac{PC}. By adjusting transmit power on each transmitted resource element individually, overall consumption is minimized.

In general, transmit power can be reduced whenever the channel conditions, in terms of the \ac{SINR}, are better than required. In addition, the existence of idle resources plays an important role. Current cellular systems and \ac{BS} are designed to provide maximum throughput and deployed such that the user demand can be fulfilled during peak hours. However, for large parts of a day, this capacity is overachieving. As a result, resource elements -- and thus parts of the available frequency band -- are unused. So instead of transmitting at high power on few resource elements, it is advantageous from an energy perspective to transmit on all available resource elements with lower transmit power. When there is more than one user requesting data, the resource elements need to be scheduled. 

\section{Minimizing transmit power}
\label{transmitpower}
\begin{figure}
\begin{center}
\includegraphics[width=0.6\textwidth]{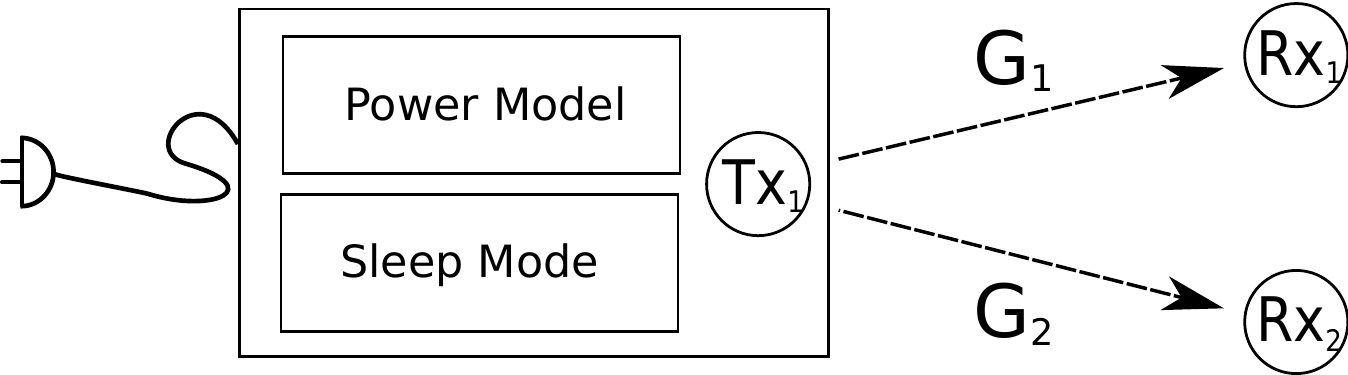}
\end{center}
\caption{The employed network model, where $G_i$ is the path gain on link $i$}
  \label{2link_MUDL}
\end{figure}

In order to describe the transmit power consumption of a \ac{BS} under varying loads, two user downlink is selected as the appropriate model, as shown in Fig.~\ref{2link_MUDL}. We proceed to derive and minimize the system transmit power consumption. According to Shannon \cite{s4801} the channel capacity of an ideal bandlimited additive white Gaussian noise channel is described by
\begin{equation}
\label{eq:Shannon}
C = W \log_2 \left( 1 + \gamma \right),
\end{equation}
where $C$ is the channel capacity in bps, $W$ is the channel bandwidth and $\gamma$ is the \ac{SINR} $\gamma = \frac{G P}{N}$ with $G$ the channel gain, $P$ the transmit power and $N$ the thermal noise.

While the bandwidth is proportional to the capacity~$C$ in \eqref{eq:Shannon}, the transmit power has a logarithmic relationship to~$C$. As a consequence, to increase data rate is much more expensive in terms of power than in terms of bandwidth. In other words, if there is a choice between a) leaving idle bands and transmitting at higher power and b) using all available bands and transmitting at the lowest required power, then option b) will always consume less overall transmit power.

The idea of a trade-off of resources against transmit power receives an additional dimension in the two-user case. There is now the option to allocate resources efficiently over two users, such that overall power consumption is minimal. To account for the spread use of the resource, we introduce the weighting factor $\mu$, representing share of the resource. In analytical terms, this is represented as follows:
\begin{equation}
C_{\rm{SYS}} = W \left[ \mu \log_2 \left( 1 + \gamma_1 \right) + (1-\mu) \log_2 \left( 1+ \gamma_2 \right) \right],
\end{equation}
where $ \gamma_1 = \frac{G_{1} P_1}{N}$, $\gamma_2 = \frac{G_{2} P_2}{N }$ on link $i$. Weighting factor $\mu = 1$ signifies the assignment of all resources to link 1, none to link 2 and vice versa for $\mu = 0$; $0 \leq \mu \leq 1$. Allocation (weighting) of resources is referred to as \ac{RS}. For a more general analysis, we consider the spectral efficiency $\varsigma$ instead of the link capacity $C$, where $\varsigma = \frac{C}{W}$.

It is our goal to identify the set of parameters that causes least transmit power consumption for a guaranteed link spectral efficiency $\varsigma_{\rm{min}}$ that is assumed to be equal on all links as the \ac{QoS} constraint:
\begin{equation}
\rm{min}~(\textit{P}_{\rm{SYS}}): \varsigma_\textit{n} \geq  \varsigma_{\rm{min}}.
\end{equation}

The overall transmit power consumed by the system is
\begin{equation}
P_{\rm{Tx}} = \mu P_1 + ( 1 - \mu ) P_2.
\label{eq:OTPSYS}
\end{equation}

The individual required link spectral efficiencies of the system on link $i$ are

\begin{equation}
\label{OTC1}
\varsigma_i = \mu \log_2\left(1+ \gamma_i \right).
\end{equation}

Thus
\begin{equation}
\label{P1}
P_1 = \left(2^{\frac{\varsigma_1}{\mu}} - 1 \right) \frac{N}{ G_{1}} \quad\rm{and}\quad P_2 = \left(2^{\frac{\varsigma_2}{1 - \mu}} - 1 \right) \frac{N}{ G_{2}}.
\end{equation}

Due to the monotonically increasing shape of the log-function, the minimum power consumption on a static link occurs for the lowest possible spectral efficiency $\varsigma_{\rm{min}}$. Therefore, the system operates optimally in terms of power consumption when
\begin{equation}
\label{C1C2Cmin}
\varsigma_1 = \varsigma_2 = \varsigma_{\rm{min}}.
\end{equation}

The combination of \eqref{eq:OTPSYS} and \eqref{P1} yields
\begin{equation}
\label{P_sysOT}
P_{\rm{Tx}} 
= \mu \frac{N}{G_{1}} \left( 2^{\frac{\varsigma_{\rm{min}}}{\mu}} - 1 \right) 
+ (1-\mu) \frac{N}{G_{2}} \left( 2^{\frac{\varsigma_{\rm{min}}}{(1 - \mu)}} - 1 \right),
\end{equation}
which describes the system transmit power consumption for a short enough time during which $G_{1}$ and $G_{2}$ remain unchanged, \ie, less than the coherence time of the channel.

For an initial analysis, Table~\ref{tab:parameterrange} lists the parameter range for the spectral efficiency~$\varsigma$, thermal noise~$N$, channel gains~$G_{i}$ and transmit powers~$P_i$, $i{=}\{1,2\}$, which corresponds to today's terrestrial cellular networks, such as \ac{LTE} or WiMAX. The noise power $N {=} -100\,$dBm represents thermal noise for $B {=} 10$\,MHz bandwidth with $N = B N_0 = B k T$, $k {=} 1.38e{-23}$\,dBm/Hz, $T {=} 290$\,K. 

\begin{table}
	\centering
		\begin{tabular}{crrl}
			\hline
			Parameter 	& Minimum   	& Maximum   	& Unit \\
			\hline
			$\varsigma$ 	& $0.1$     	& $6$       	& bps/Hz \\
			$N$        	& $-100$    	& $-100$	& dBm \\
			$G_{i}$   	& $-150$	& $-50$		& dB  \\
			$P_i$      	& $-50$		& $50$		& dBm \\
			\hline
		\end{tabular}
	\caption{Evaluated parameter ranges}
	\label{tab:parameterrange}
\end{table}

\begin{figure}
\centering
\includegraphics[width=0.5\textwidth]{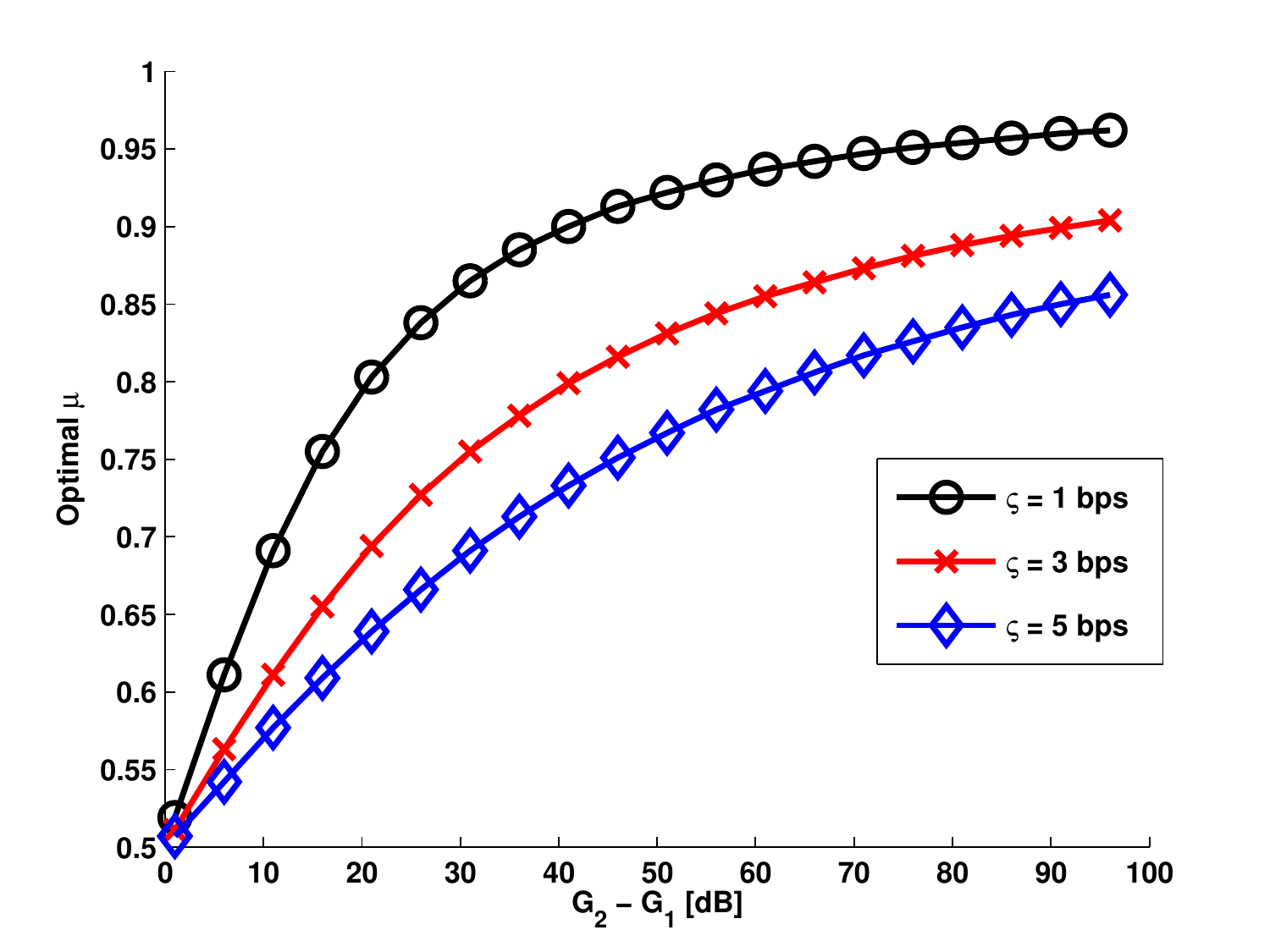}
\caption{Optimal weighting factor $\mu$ under selected spectral efficiencies.}
\label{fig:optimalMu2D}
\end{figure}

For each combination of parameters, there is one optimal weighting factor $\mu$ that minimizes power consumption. Figure~\ref{fig:optimalMu2D} shows the optimal weighting factor for selected spectral efficiencies. For high required spectral efficiencies, $\mu$ approaches linearity.

\section{From Transmit Power to Supply Power}
\label{supplypower}
While the findings from the previous section establish a necessary foundation for energy-efficient transmission, there is one more step necessary to generate a complete picture. So far, only transmit power has been considered. By means of a power model, which translates transmit power into supply power (``power at the 230V plug``) of a \ac{BS}, it is possible to account for component behavior. While the model has recently been described to behave linearly \cite{czbfjgav1001}, it still adds one important dimension to the problem: A \ac{BS} consumes significant power even when it is not transmitting. For transmit power optimization this bears the significant consequence that savings at low transmit powers have no significant effect on the supply power side. In low power/low load operation, the standby consumption causes the bulk of overall consumption. Instead an intermediate step can be taken via sleep mode. This means that the \ac{BS} enters a micro sleep as soon as it is not needed. There exist several levels or ``depths`` of sleep which take different amounts of time to enter and leave. For this study, we assume that there exists exactly one sleep mode that is available instantly. In the following, the problem presented in the previous section is extended by two additional steps. First, transmit power is translated into supply power via the linear power model. Second, there exists the alternative for the \ac{BS} to be in sleep mode during which it consumes a lower-than-standby amount of power. Depending on the share of time it spends in sleep mode, the standby consumption is reduced. However, as a tradeoff, the rate has to be increased during awake-time. This in turn causes higher necessary transmit power.

Formally, the power model effect is added by 
\begin{equation}
 P_{\rm{PM}} = P_0 + m P_{\rm{Tx}} 
\end{equation}
where $P_0$ is the standby power consumption and $m$ is the slope of the load dependent power consumption. 

To account for the effect of \ac{DTX}, the activity factor $t$ is introduced which indicates the share of time spent in active mode (rather than sleep mode). \ac{DTX} power is modeled as
\begin{equation}
 P_{\rm{supply}} = ( 1 - t ) P_{\rm{s}} + t P_{\rm{PM}},
\end{equation}
where $t$ is the share of time spent in active (awake) mode and $P_s$ is the \ac{BS} power consumption in sleep mode.

The trade-off of \ac{DTX} is an increased rate during active time. This is included in the model via $\varsigma_{\rm{DTX}} = \frac{\varsigma}{t}$.

As has been shown so far, minimum transmit power leads to minimum power consumption under the linear power model. However, additional savings are possible by employing \ac{DTX}, which completely changes the transmit-power-only analysis for low data rates. 

\section{Results and Discussion}
\label{results}
\begin{figure}
\centering
\includegraphics[width=0.7\textwidth]{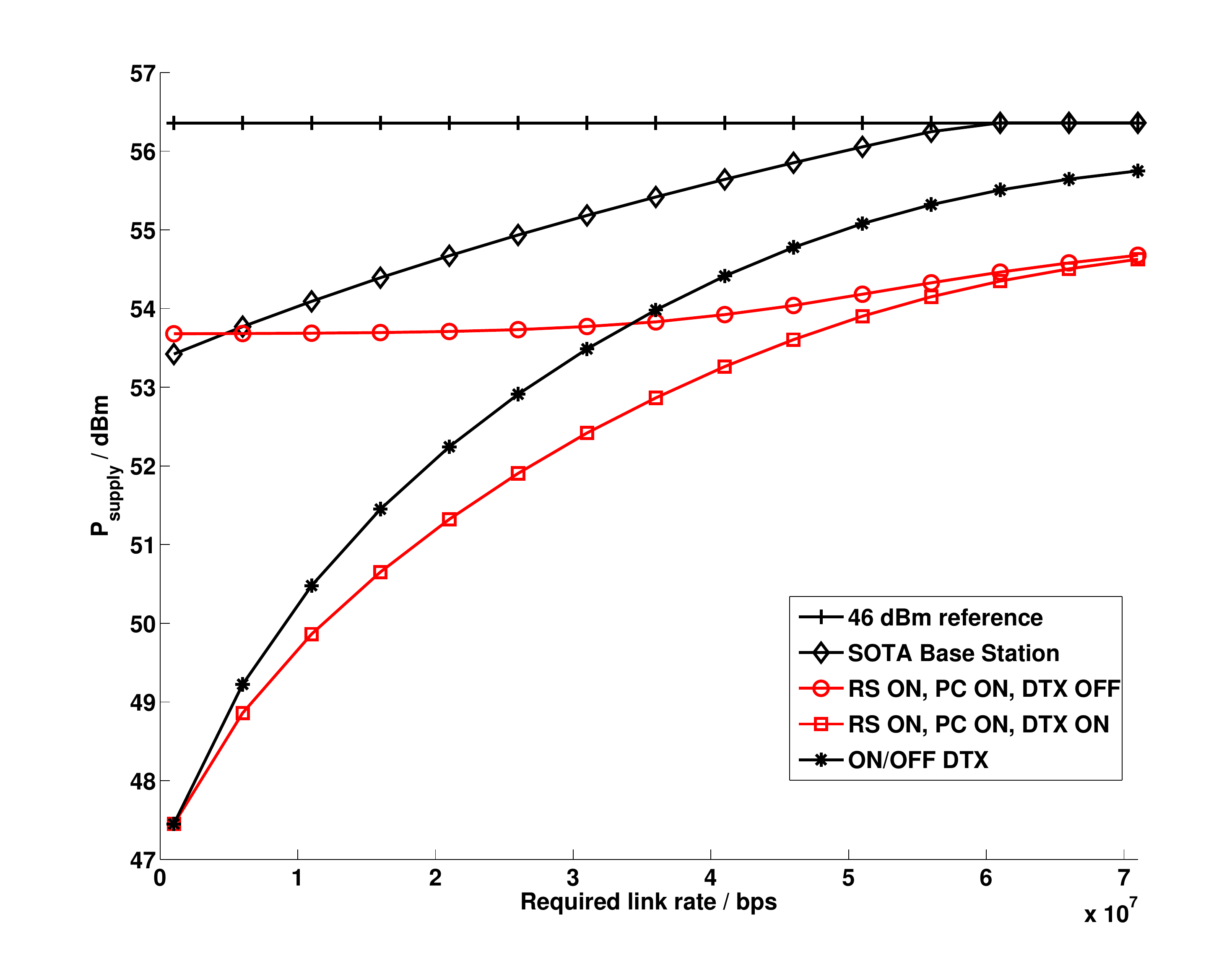}
\caption{Comparison of supply power consumption as a function of link rate under varying schemes}
\label{fig:PsysSchemeComparison}
\end{figure}

The described two-user downlink system has been evaluated in simulation using the system parameters listed in Table~\ref{tab:uniformdistrscenarioparameters}. Two users are dropped uniformly on a disk, generating a distribution of channel gains. Results of the achievable data rate are plotted over the bandwidth. Power consumption is plotted for the power-optimal selection of weighting factor~$\mu$ and awake time share~$t$ in Figure~\ref{fig:PsysSchemeComparison}.

    \begin{table}
\centering
      \begin{tabular}{c|c}
	Parameter          			& Value\\
	\hline
	Carrier frequency  			& 2 GHz\\
	Cell radius  				& 250 m\\
	Pathloss model  			& 3GPP UMa, NLOS, shadowing\\
	Shadowing standard deviation		& 8 dB\\
	Iterations  				& 10,000 \\
	Bandwidth  				& 10 MHz \\
	Noise power  				& -100 dBm \\	
	$P_0$, \ac{BS} power consumption at zero load  			& 233 W\\
	$m$, Increase per Watt PTx (model slope)  			& 5\\
	$P_{\rm{s}}$, \ac{BS} power consumption in sleep mode  		& 50 W\\
	Pmax  								& 46 dBm \\
	\end{tabular}
      \caption{Simulation Parameters}
\label{tab:uniformdistrscenarioparameters}
    \end{table}

While the \ac{RS}-\ac{PC}-\ac{DTX} scheme provides power-optimal resource allocation, it is necessary to define alternative allocation schemes for reference and estimation of achievable gains. For comparison, the following schemes are included in the analysis: A constant transmit power of 46~dBm which represents the upper limit of an \ac{LTE} macro \ac{BS}; a weakly load dependent \ac{SOTA} \ac{BS} as the reference; \ac{RS}-\ac{PC} without \ac{DTX}; a scheme without power control, but with \ac{DTX}; and the optimal \ac{RS}-\ac{PC}-\ac{DTX} scheme. The 46\,dBm transmit power reference provides an absolute upper limit, which represents the supply power consumption of a \ac{BS} that is always-on and transmitting at maximum power. However, this is not a realistic assumption since even \ac{BS} today (\ac{SOTA}) that are not manufactured for energy efficiency have some load dependent behavior. As an estimate, we assume that such a \ac{SOTA} \ac{BS} consumes only half of its maximum consumption (\ie, $-3$\,dB) when it is not loaded with a linear raise in consumption up to full load \cite{czbfjgav1001}. The first allocation scheme is one where allocation of resource blocks and \ac{PC} are enabled, but the \ac{BS} does not have the capability of sleep modes. This is informative to estimate gains in addition to \ac{DTX}. In contrast, it is also instructive to consider a scheme which does not have \ac{PC} available, but can go into sleep mode, which is labelled 'ON/OFF DTX'. Ultimately, the optimal \ac{RS}-\ac{PC}-\ac{DTX} allocation scheme utilizes all three available options. 

Figure~\ref{fig:PsysSchemeComparison} summarizes our findings. The overall system power consumption $P_{\rm{supply}}$ is plotted vs. the required link rate. It is important to note that only link rates up to around $6\times10^7$~bps can be provided without outage. Thus, the plot values for link rates higher than $6\times10^7$~bps are provided as a trend. For the same reason, the SOTA \ac{BS} consumption is only plotted over this range.

Most relevant are the achievable gains of the optimal scheme (\ac{RS}-\ac{PC}-\ac{DTX}) compared to the \ac{SOTA} behavior. Here, it can be seen that the largest gains are expected in low load (5.5~dB) and slowly decrease (to 2~dB) as the required link rate approaches $6\times10^7$\,bps. In percentages, these fundamental limits translate into a 73\% savings potential in low load going towards a 27\% saving in high load situations. In addition to these cornerstone numbers, further relevant behavior can be identified. Comparing 'ON/OFF DTX' with the optimal scheme shows that in low load \ac{PC} has hardly an effect. This is due to standby power consumption outmatching the transmit power consumption by orders of magnitude in the power model. In other words, saving transmit power in a low load situation bears insignificant benefit. Only as load increases (higher required rates) to the curves part, yielding a 1\,dB benefit of the optimal scheme over 'ON/OFF DTX'. This benefit comes at the cost of allowing downlink power control which stretches today's standards specification and technical capabilities. The merging of the two \ac{RS}-\ac{PC} curves illustrates that --- as rates grow --- \ac{DTX} becomes less feasible. In this operating region, situations occur in which it is more efficient to keep the \ac{BS} active (\ie, no sleep mode) and decrease transmit power than to go into sleep mode.

\section{Conclusions}

This study reveals that \ac{DTX} is needed in future energy-efficient \ac{BS}, if significant gains are expected. Power control alone can only reduce power consumption by around 20\% in high load situations. When optimizing for low load, \ac{DTX} by itself delivers similar gains as a scheme which employs \ac{RS}, \ac{PC} and \ac{DTX}. However, only the combination of all three techniques is overall power-optimal.

When considering today's hardware and standards definitions, \ac{DTX} is the appropriate energy saving scheme in range. However, consideration of the control signalling in \ac{LTE} is part of future work in this field and is expected to reduce the applicability of \ac{DTX}, because during necessary control signalling the \ac{BS} is not allowed to enter sleep mode.

\section*{Acknowledgements}

 The authors gratefully acknowledge the contribution of the \ac{EARTH} consortium to the technical definition of the project.

 The work leading to this paper has received funding from the European Community's Seventh Framework Programme [FP7/2007-2013. EARTH, Energy Aware Radio and neTwork tecHnologies] under grant agreement n${}^\circ$ 247733.


\bibliography{../../../../../../DOCOMO/reference/DOCOMO,../../../../../../DOCOMO/reference/general}
\bibliographystyle{ieeetr}

\end{document}